\documentclass[10pt]{article}
\usepackage{graphicx}
\usepackage{xspace}
\usepackage[bookmarks=true,linktocpage]{hyperref}
\usepackage{cite}
\hypersetup{%
  pdftitle={SUSY Global Fits},
  pdfauthor={Peter Athron},
  colorlinks=true,urlcolor=blue,linkcolor=blue,citecolor=blue,filecolor=blue}

\newcommand{\GB}{\textsf{GAMBIT}\xspace}
\newcommand{\Diver}{\textsf{Diver}\xspace}
\newcommand{\MultiNest}{\textsf{MultiNest}\xspace}
\newcommand{\flexiblesusy}{\textsf{FlexibleSUSY}\xspace}
\newcommand{\softsusy}{\textsf{SOFTSUSY}\xspace}
\newcommand{\sarah}{\textsf{SARAH}\xspace}
\newcommand{\HDECAY}{\textsf{HDECAY}\xspace}
\newcommand{\SDECAY}{\textsf{SDECAY}\xspace}
\newcommand{\SUSYHIT}{\textsf{SUSYHIT}\xspace}
\newcommand{\GMtwoCalc}{\textsf{GM2Calc}\xspace}
\newcommand{\HiggsBounds}{\textsf{HiggsBounds}\xspace}
\newcommand{\HiggsSignals}{\textsf{HiggsSignals}\xspace}
\newcommand{\SuperIso}{\textsf{SuperIso}\xspace}
\newcommand{\DarkSUSY}{\textsf{DarkSUSY}\xspace}
\newcommand{\DDCalc}{\textsf{DDCalc}\xspace}
\newcommand{\Pythia}{\textsf{Pythia}\xspace}
\newcommand{\BuckFast}{\textsf{BuckFast}\xspace}
\newcommand{\microOMEGAs}{\textsf{MicroOMEGAs}\xspace}
\newcommand{\nuLike}{\textsf{nulike}\xspace}
\def\Title#1{\begin{center} {\Large #1 } \end{center}}
\def\Author#1{\begin{center}{ \sc #1} \end{center}}
\def\Address#1{\begin{center}{ \it #1} \end{center}}

\newcommand\pubblock{\rightline{\begin{tabular}{l} Proceedings of the Fifth Annual LHCP\\ \pubnumber\\
         \pubdate  \end{tabular}}}

\newenvironment{Abstract}{\begin{quotation} \begin{center}
      \large \end{center}\bigskip
      \begin{center}\begin{large}}{\end{large}\end{center} \end{quotation}}

\newenvironment{Presented}{\begin{quotation} \begin{center} 
             PRESENTED AT\end{center}\bigskip 
      \begin{center}\begin{large}}{\end{large}\end{center} \end{quotation}}

\def\Acknowledgements{\bigskip  \bigskip \begin{center} \begin{large}
             \bf ACKNOWLEDGEMENTS \end{large}\end{center}}

\def\beq{\begin{equation}}
\def\eeq#1{\label{#1}\end{equation}}
\def\eeqn{\end{equation}}

\def\beqa{\begin{eqnarray}}
\def\eeqa#1{\label{#1}\end{eqnarray}}
\def\eeqan{\end{eqnarray}}

\let\bar=\overbar

\def\Dslash{\not{\hbox{\kern-4pt $D$}}}
\def\dslash{\not{\hbox{\kern-2pt $\del$}}}

\def\msb{{\bar{\ssstyle M \kern -1pt S}}}

\usepackage{color}

\newcommand\pubnumber{ CoEPP--MN--17--14 }

\newcommand\pubdate{\today}

\def\affiliation{
On behalf of the GAMBIT collaboration, \\
School of Physics and Astronomy \\
Monash University, Melbourne, VIC 3800, Australia}

\begin{document}

\large
\begin{titlepage}
\pubblock

\vfill
\Title{ SUSY Global Fits  }
\vfill

\Author{ Peter Athron  }
\Address{\affiliation}
\vfill
\begin{Abstract}
\begin{minipage}{14cm}  
We present comprehensive global fits of supersymmetric (SUSY) models
from the Global and Modular Beyond-the-Standard-Model Inference Tool
(\GB) collaboration, based on arXiv:1705.07935 and arXiv:1705.07917.
We investigate several variants of the minimal supersymmetric standard
model, a fully constrained version (CMSSM) with universal scalar
($m_0$), gaugino ($m_{1/2}$) and trilinear masses ($A_0$) at the gauge
coupling unification scale, a similar model that is relaxed by adding
an extra parameter for the soft Higgs masses (NUHM1), another where
the soft Higgs masses are also split (NUHM2) and finally a weak scale
MSSM7 model. We use the public \GB global fitting framework and take
into account all relevant data to reveal the regions of parameter
space with the highest likelihood. Our results reveal that all models
have very heavy scenarios that are well out of reach of the LHC, but
will be probed by forthcoming dark matter experiments, as well as a
stop-co-annihilation region which has better prospects for detection
in collider experiments.  The stau co-annihilation region is excluded
from the CMSSM at $2 \sigma$ but is present in the NUHM1 and NUHM2
variants. Finally by relaxing constraints in the NUHM1, NUHM2 and
MSSM7 we see additional regions appear: lighter chargino
co-annihilation region, sbottom co-annihilation and $h/Z$ funnels.
\end{minipage}



\end{Abstract}
\vfill

\begin{Presented}
The Fifth Annual Conference\\
 on Large Hadron Collider Physics \\
Shanghai Jiao Tong University, Shanghai, China\\ 
May 15-20, 2017
\end{Presented}
\vfill
\end{titlepage}
\def\thefootnote{\fnsymbol{footnote}}
\setcounter{footnote}{0}
%

\normalsize 


\section{Introduction}
All realistic SUSY models have both a large multidimensional parameter
space and many collider and astrophysical observables can constrain
that parameter space.  To understand the combined impact of all of
these observables on the model we need to perform global fits of these
models using rigorous statistical methods and intelligent scanning
algorithms. To do this we have created \GB, which is both a new global
fitting collaboration and a public tool for performing global fits of
models of beyond the standard model physics \cite{Athron:2017ard,
  Workgroup:2017htr, Workgroup:2017bkh, Workgroup:2017lvb,
  Balazs:2017moi, Workgroup:2017myk}.  \GB is designed to work with
any standard model (SM) extension (both non-SUSY and non-minimal
SUSY), but in this proceedings we focus\footnote{For a \GB global fit
  with a non-SUSY model we refer the reader to the global fit of the
  scalar singlet model with \GB \cite{Athron:2017kgt} that was also
  released at the same time as the SUSY papers, but will not be
  discussed here.} on the results in the minimal supersymmetric
standard model (MSSM). We present results in the CMSSM, NUHM1 and
NUHM2 based on work carried out in Ref.\ \cite{Athron:2017qdc} and
also for a weak scale MSSM7 from Ref.\ \cite{Athron:2017yua}.

\GB allows the user to choose between a number of different calculators for the MSSM. For the results discussed here we used the following backends (external codes): \Diver \cite{Workgroup:2017htr} and \MultiNest 3.10 \cite{Feroz:2008xx} (efficient sampling); \flexiblesusy 1.5.1\footnote{\flexiblesusy uses \sarah 4.9.1 \cite{Staub:2009bi, Staub:2010jh,Staub:2012pb,Staub:2013tta} and contains some numerical routines from \softsusy \cite{Allanach:2001kg,Allanach:2013kza}} \cite{Athron:2014yba} (mass spectrum generation); \HDECAY \cite{Djouadi:1997yw} and \SDECAY \cite{Muhlleitner:2003vg}  in \SUSYHIT 1.5 \cite{Djouadi:2006bz} (Decay BRs and widths); \HiggsBounds 4.3.1 \cite{Bechtle:2008jh} and \HiggsSignals 1.4.0 \cite{Bechtle:2013xfa} (Higgs likelihoods); \Pythia-8.212 \cite{Sjostrand:2014zea} and \BuckFast \cite{Balazs:2017moi} (used by the native recast tool in ColliderBit \cite{Balazs:2017moi}); \GMtwoCalc 1.3.1 \cite{Athron:2015rva} (anomalous magnetic moment of the muon); \SuperIso 3.6 \cite{Mahmoudi:2007vz} (flavour physics); \microOMEGAs 3.6.9.2 \cite{Belanger:2001fz}, \DarkSUSY 5.1.3 \cite{Gondolo:2004sc}, \DDCalc 1.0.0 \cite{Workgroup:2017lvb} \nuLike 1.0.0 \cite{Scott:2012mq} (dark matter relic density and direct and indirect detection).

\section{Results}
\begin{figure}[htb]
\centering
\includegraphics[height=2.5in]{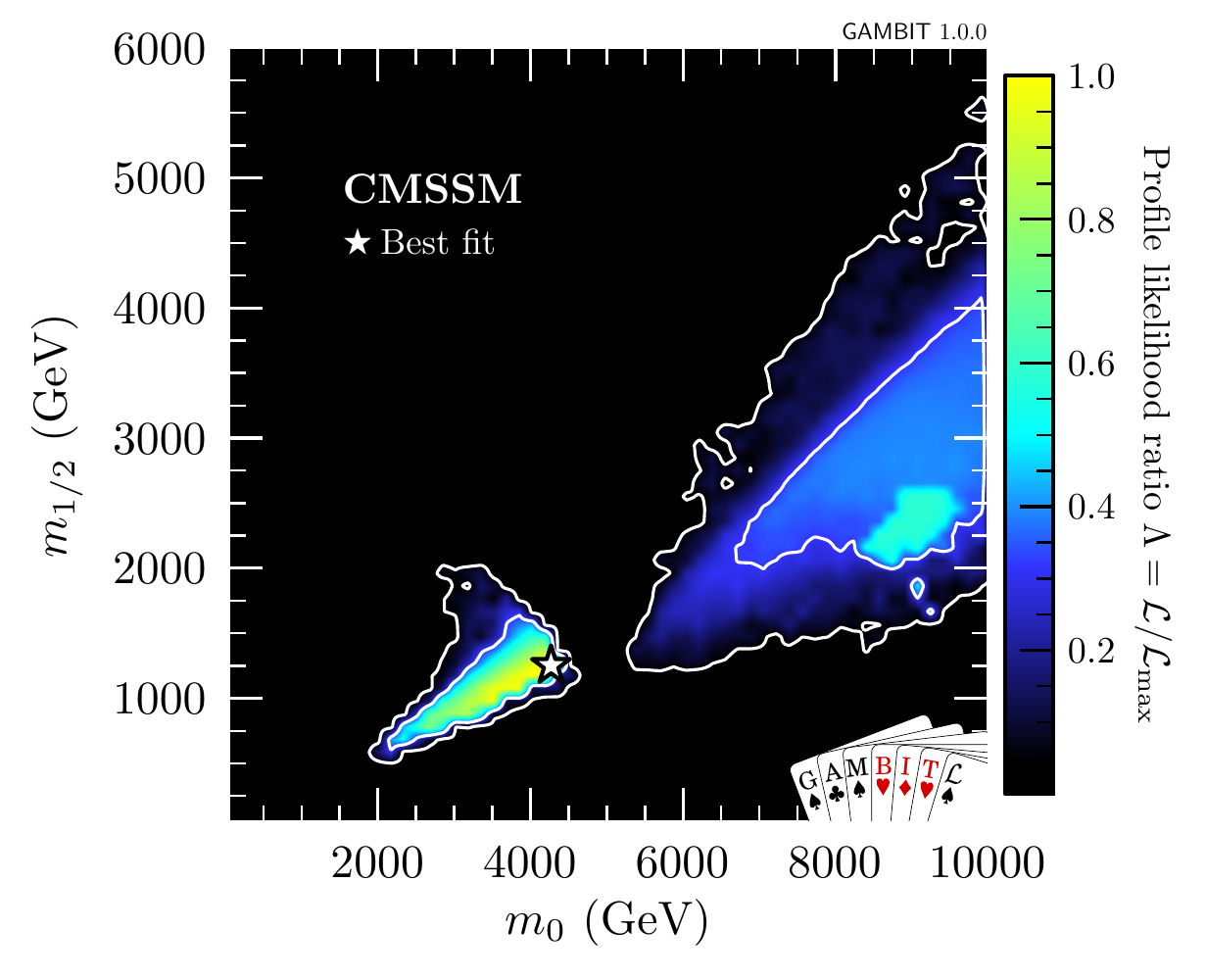}
\includegraphics[height=2.5in]{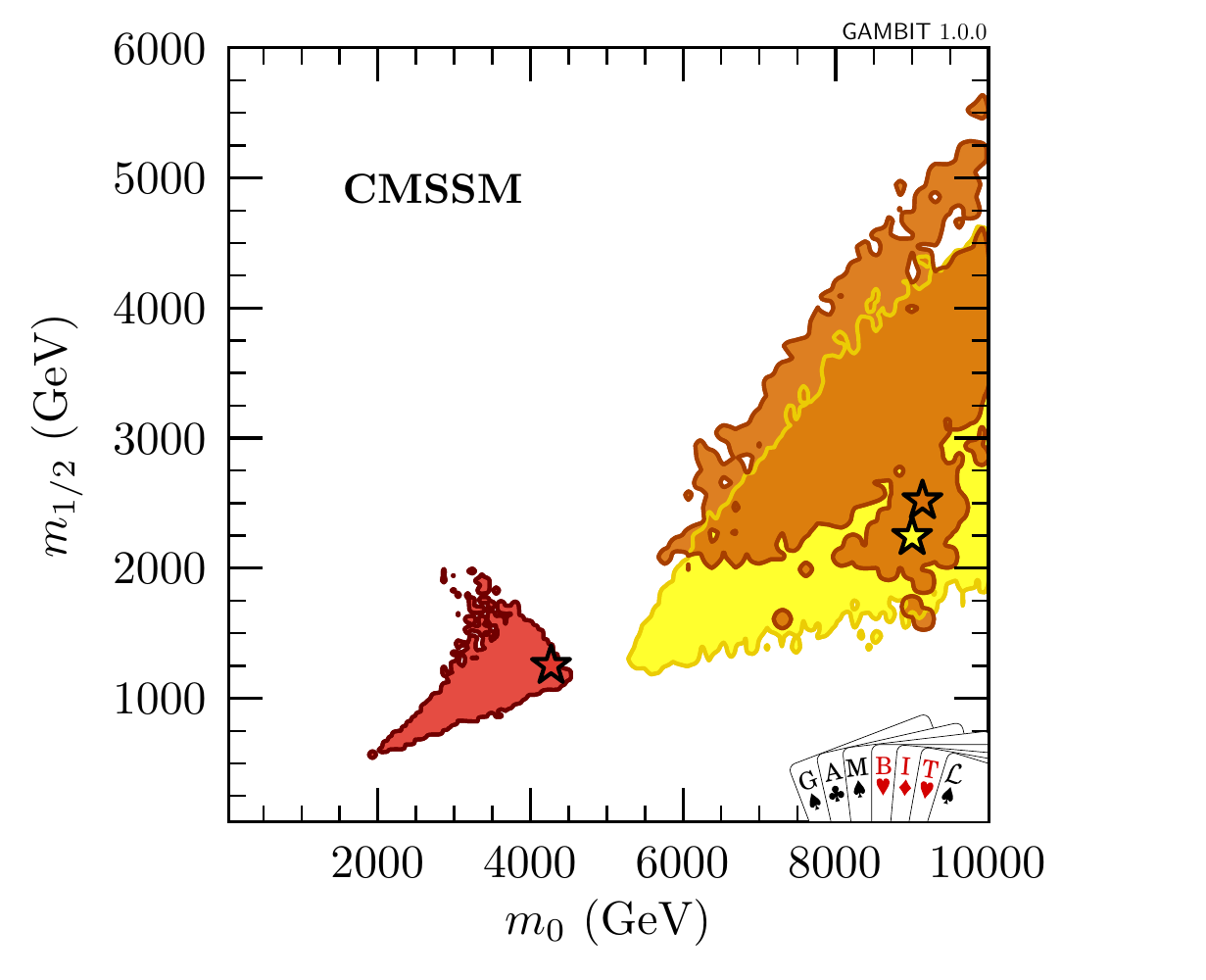}
\caption{CMSSM results in the $m_0-m_{1/2}$ plane with $2 \sigma$ confidence regions. The profile likelihood ratio is shown as the colour contour (left panel) and the active annihilation mechanisms to avoid overproduction of neutralino dark matter are indicated by colour coding: stop co-annihilation (red), chargino co-annihilation (yellow), $A$-funnel (brown).  Note that we show all active mechanism and colors overlap.}
\label{fig:CMSSMm0m12}
\end{figure}
 The results for the CMSSM global fit are shown in Fig.\ \ref{fig:CMSSMm0m12} as $2 \sigma$ confidence regions in the $m_0-m_{1/2}$ plane.  We see a large region at high $m_0-m_{1/2}$ with sparticles that are well out of reach of the LHC.  This region has $A$-funnel and chargino co-annihilation regions active. We also see a stop co-annihilation region which did not appear in previous global fits\footnote{It was also found in the very recent results of Ref.\ \cite{Han:2016gvr}. } and this region contains our best fit point.  In contrast to previous fits, we find that the stau co-annihilation region is now excluded from the $2 \sigma$ confidence regions. 

\begin{figure}[htb]
\centering
\includegraphics[height=2.5in]{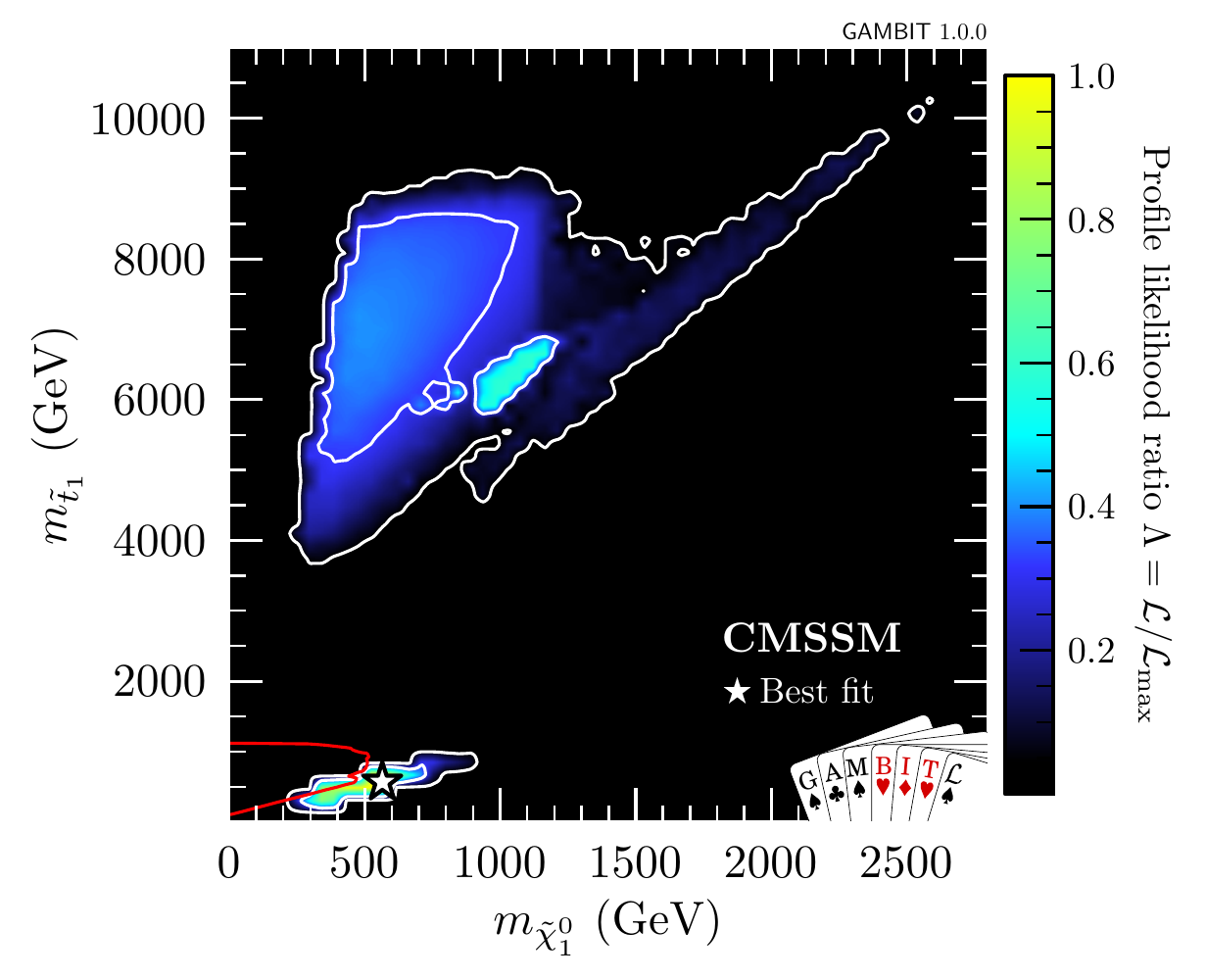}
\includegraphics[height=2.5in]{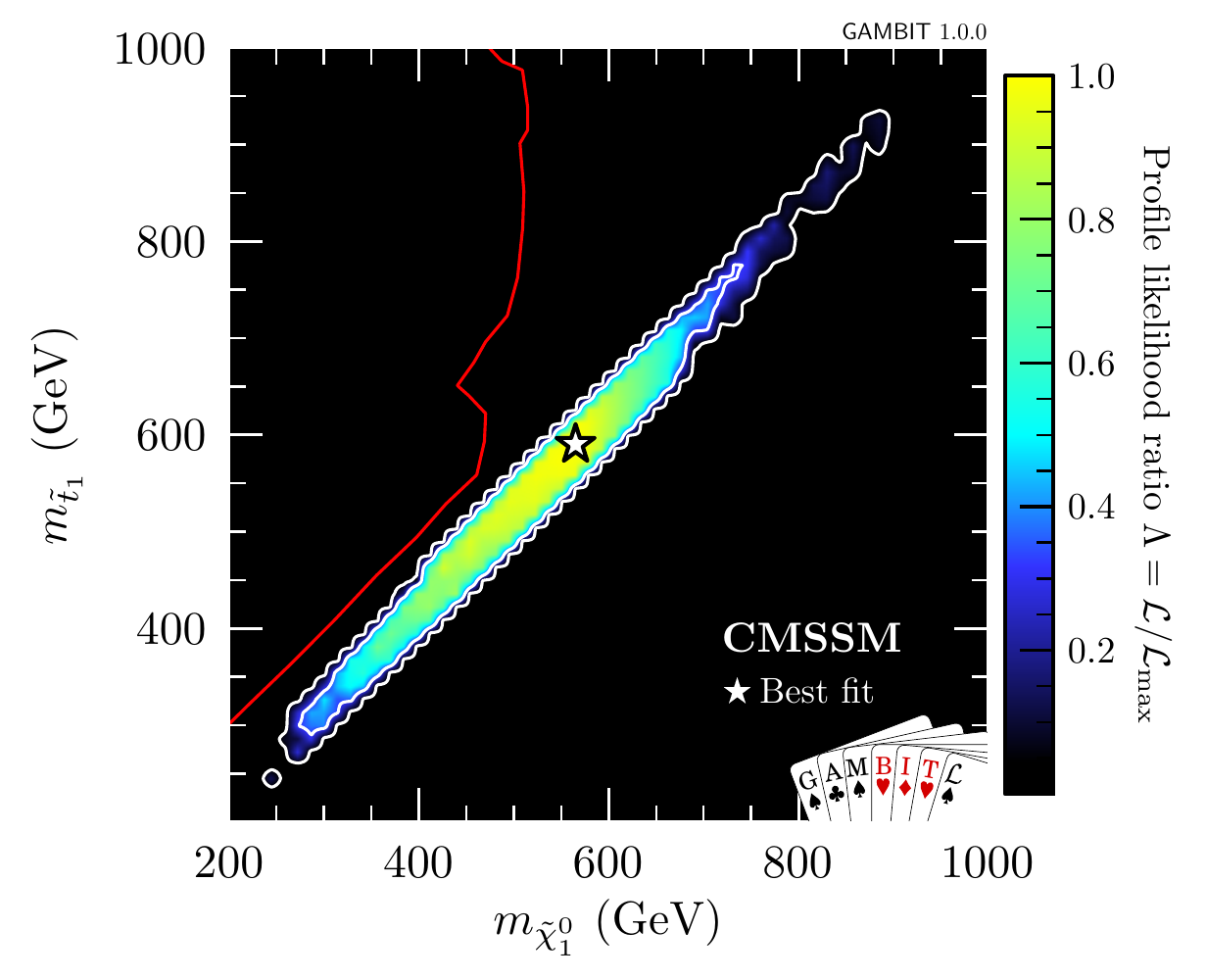}
\caption{ CMSSM results in the $m_{\tilde{t}_1} - m_{\chi_1^0}$ plane with the profile likelihood ratio shown in colour contour.  The full $2 \sigma$ confidence regions are shown in the left plot while the right plot zooms in on the stop co-annihilation region.  The red line indicates current limits from CMS compressed spectra searches which were not included in the fit.}
\label{fig:CMSSMStopCoan}
\end{figure}

Fig.\ \ref{fig:CMSSMStopCoan} shows our results in the
$m_{\tilde{t}_1} - m_{\chi_1^0}$ plane. From this one can see the stop
and lightest neutralino masses which are most important for assessing
the impact of current and future long lived sparticle or compressed
spectra searches for the light stop in the stop co-annihilation
region.  We overlayed the plot with a red line showing the limit from
current CMS compressed spectra searches, indicating that this region
is currently not constrained by these but that there is an opportunity
for future searches to have an impact here.  It is also important to
note that the large trilinear couplings needed to obtain this region
could potentially lead to vacuum instabilities and this requires
careful study on the theory, involving precise calculation of the
Higgs mass and the minima of the potential and phase transition
properties.

\begin{figure}[htb]
\centering
\includegraphics[height=2.5in]{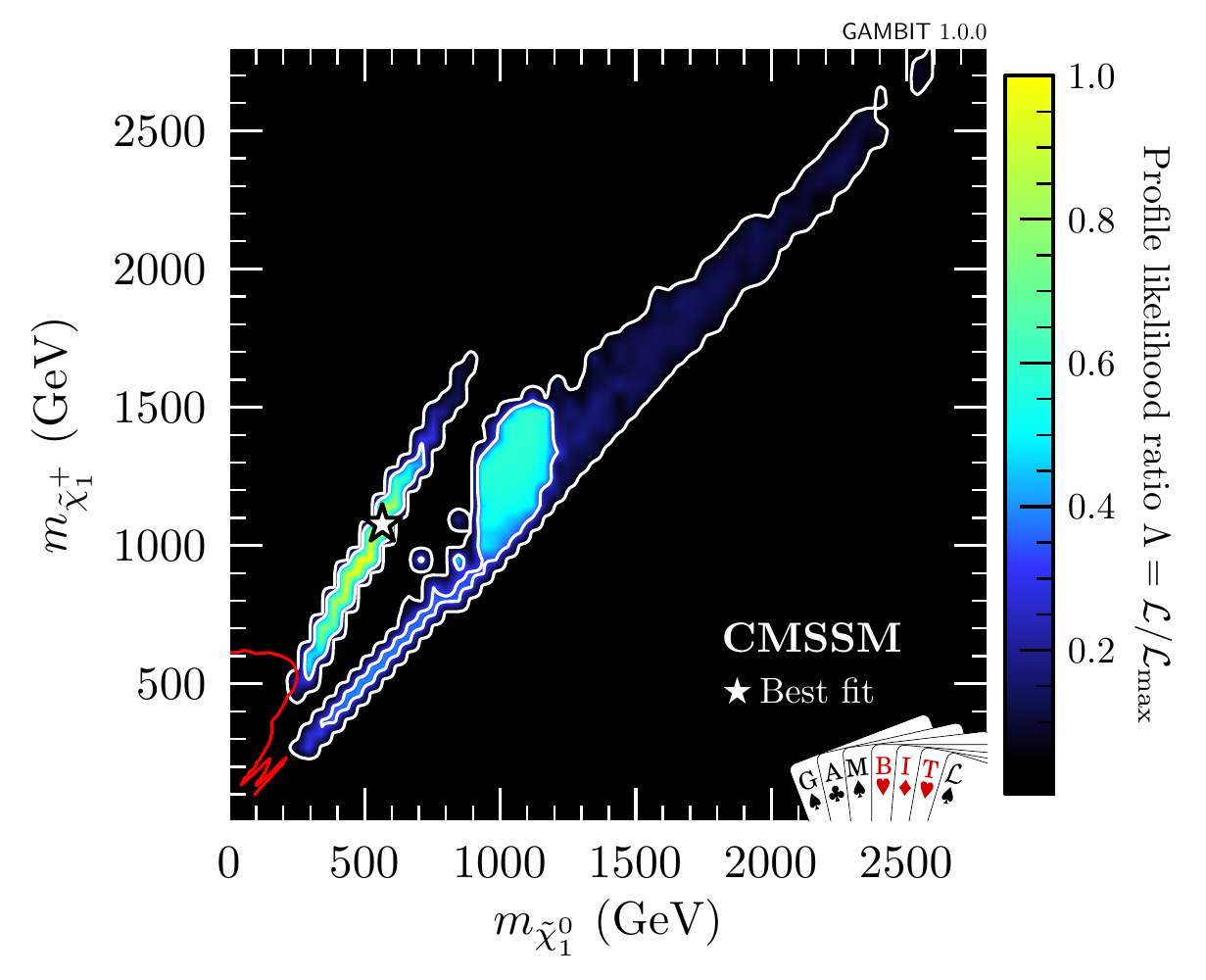}
\includegraphics[height=2.5in]{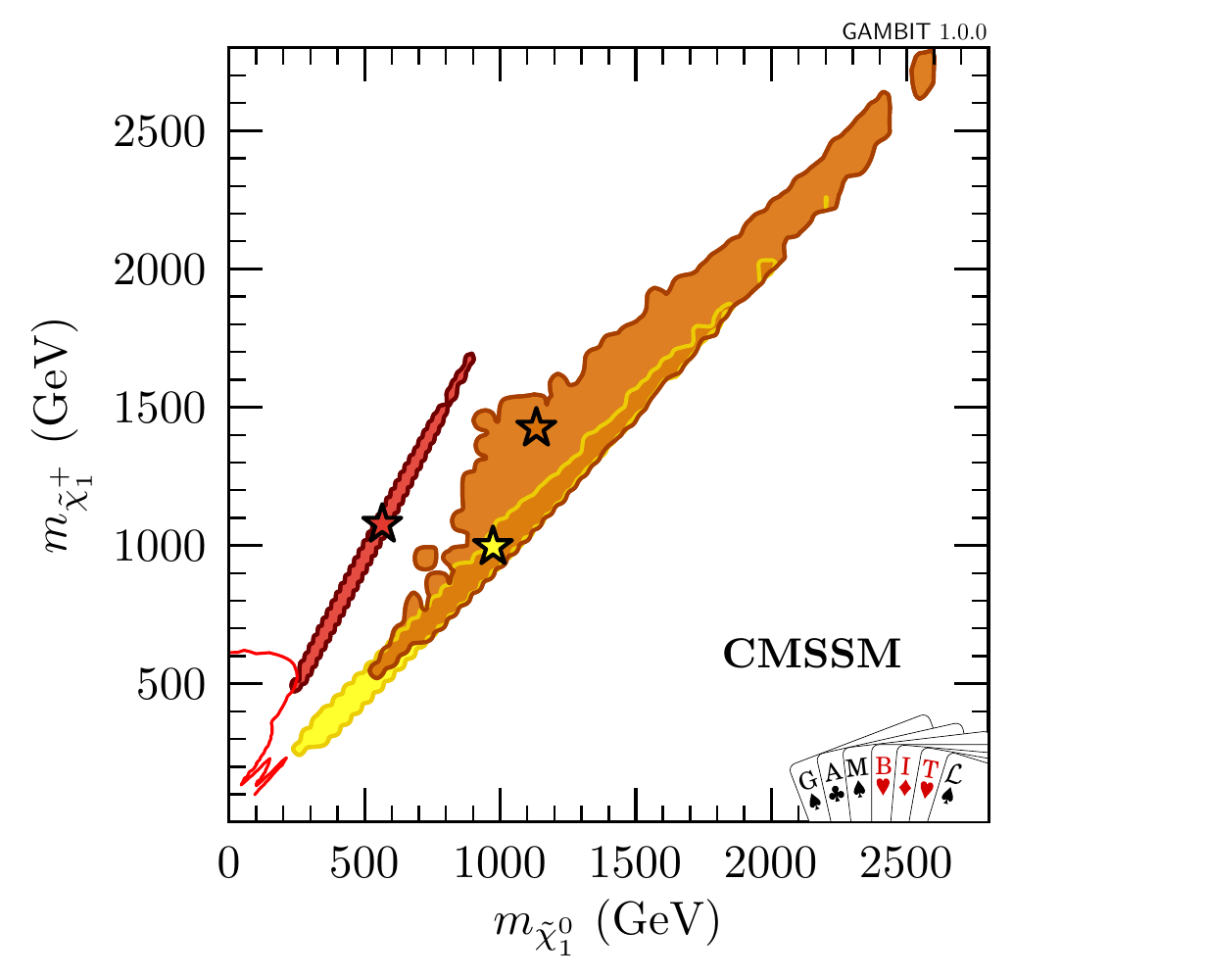}
\caption{CMSSM results in $m_{\chi_1^0}-m_{\chi_1^\pm}$ plane with profile likelihood shown as a colour contour and relic density mechanism indicated by colour as in Fig.\ \ref{fig:CMSSMm0m12}. }
\label{fig:CMSSMcharginos}
\end{figure}

In Fig.\ \ref{fig:CMSSMcharginos} we also see that pair production of
the light bino and wino associated with the stop co-annihilation
region could be detected in future, with the latest CMS run II
searches for this from a simplified model interpretation (red line)
just touching this region.  At the same time we also see a light
chargino-co-annihilation region with light charged and neutral
Higgsinos, which is only allowed because we apply the relic density
measurement as an upper limit only and allow under abundance of the
relic density.  However while it is possible to obtain such light states in
the CMSSM they are still very challenging to detect at the LHC because
of the very small mass splitting between the charged and neutral
Higgsinos.

\begin{figure}[htb]
\centering
\includegraphics[height=2.5in]{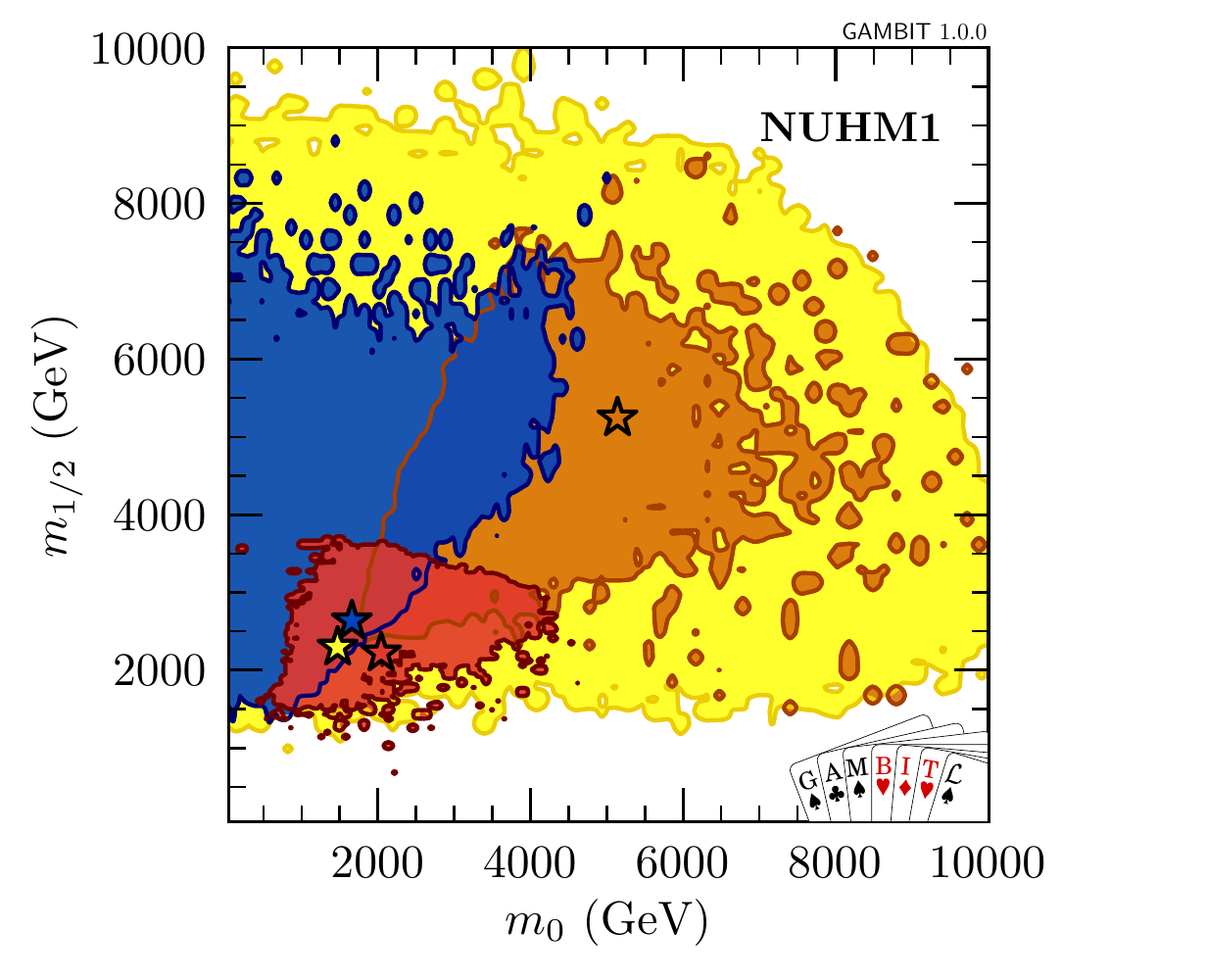}
\includegraphics[height=2.5in]{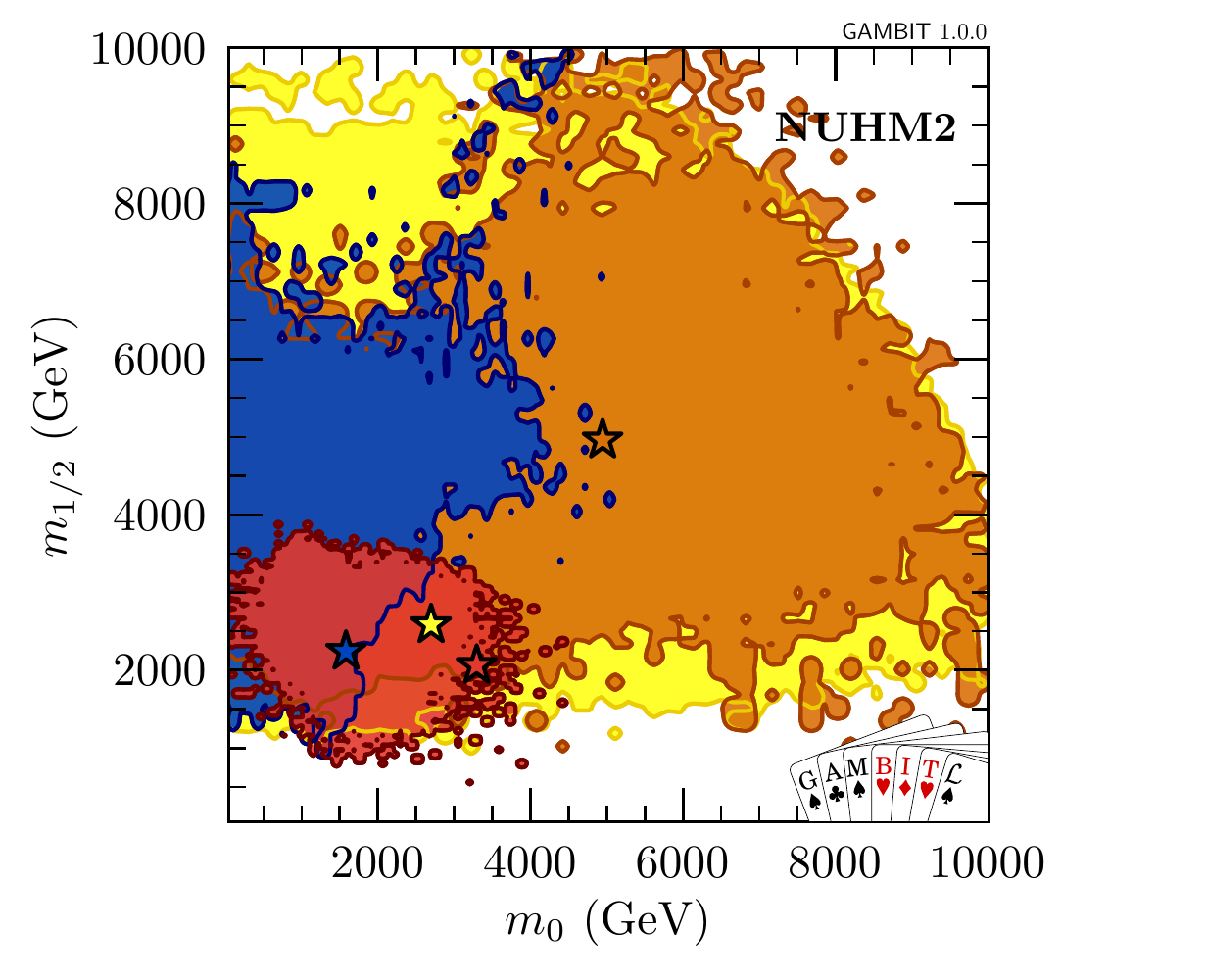}
\caption{ NUHM results in the $m_0-m_{1/2}$ plane with $2 \sigma$
  confidence regions. The profile likelihood ratio is shown as the
  colour contour (left panel) and as for Fig.\ \ref{fig:CMSSMm0m12}
  annihilation mechanism are indicated with the new stau-co-annihilation
  region shown as blue.}
\label{fig:NUHM}
\end{figure}

Results in the NUHM1 and NUHM2 are shown in Fig.\ \ref{fig:NUHM} in
the $m_0-m_{1/2}$ plane. These results include lighter first and
second generation squarks than can be obtained in the CMSSM, but these
are still out of reach for the LHC experiments. The
stau-co-annihilation region re-emerges in these less constrained
models, however the light stau associated with this is still too heavy
to be seen at the LHC.  In these models some of the stop
co-annihilation region may already be excluded by searches for
compressed spectra, but it also extends to heavier masses that are
very challenging to exclude. As in the CMSSM there are also large
regions of chargino co-annihilation and $A$-funnel with very heavy
spectra that are well out reach of the LHC.

\begin{figure}[htb]
  \centering
  \includegraphics[height=2.5in]{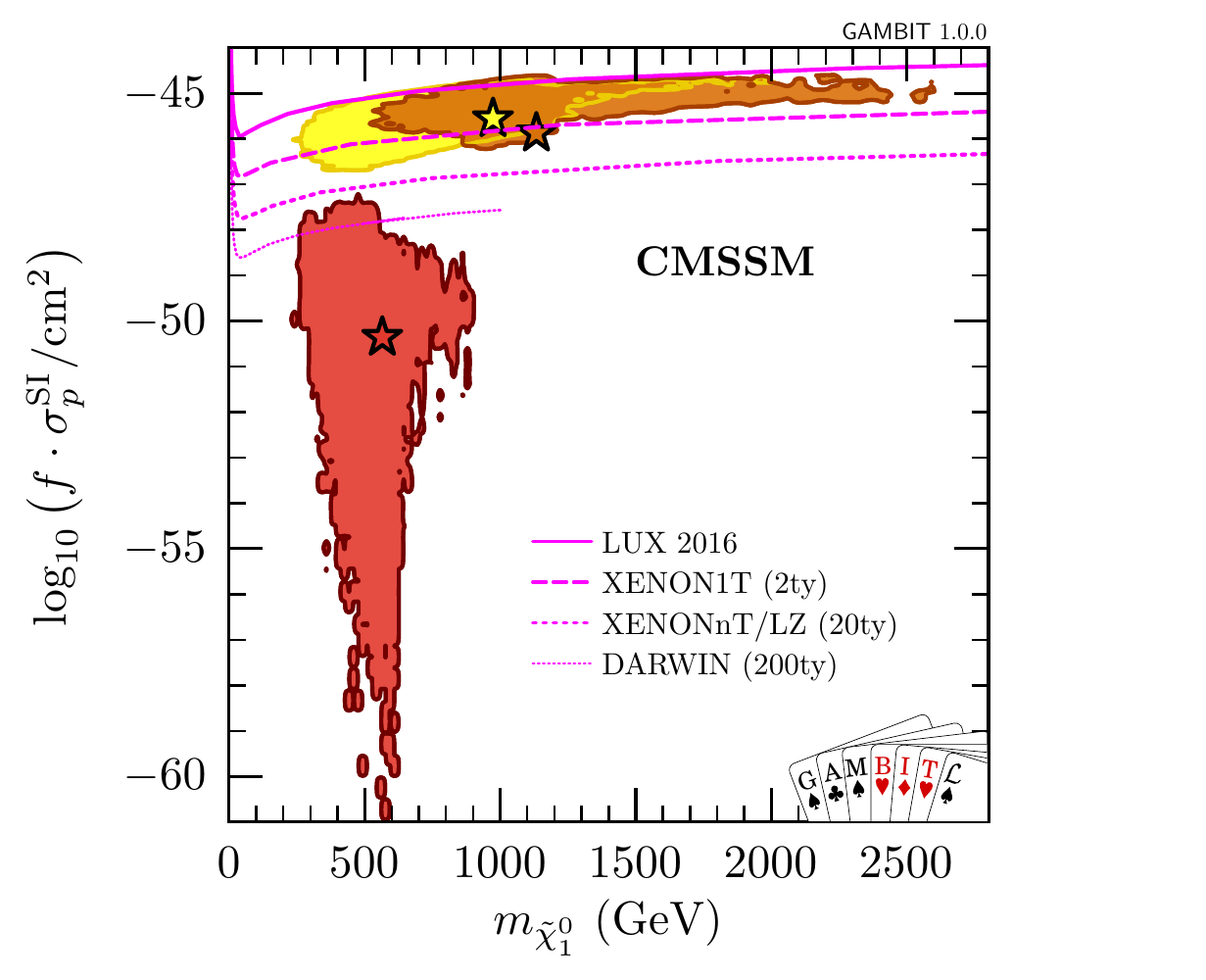}
  \includegraphics[height=2.5in]{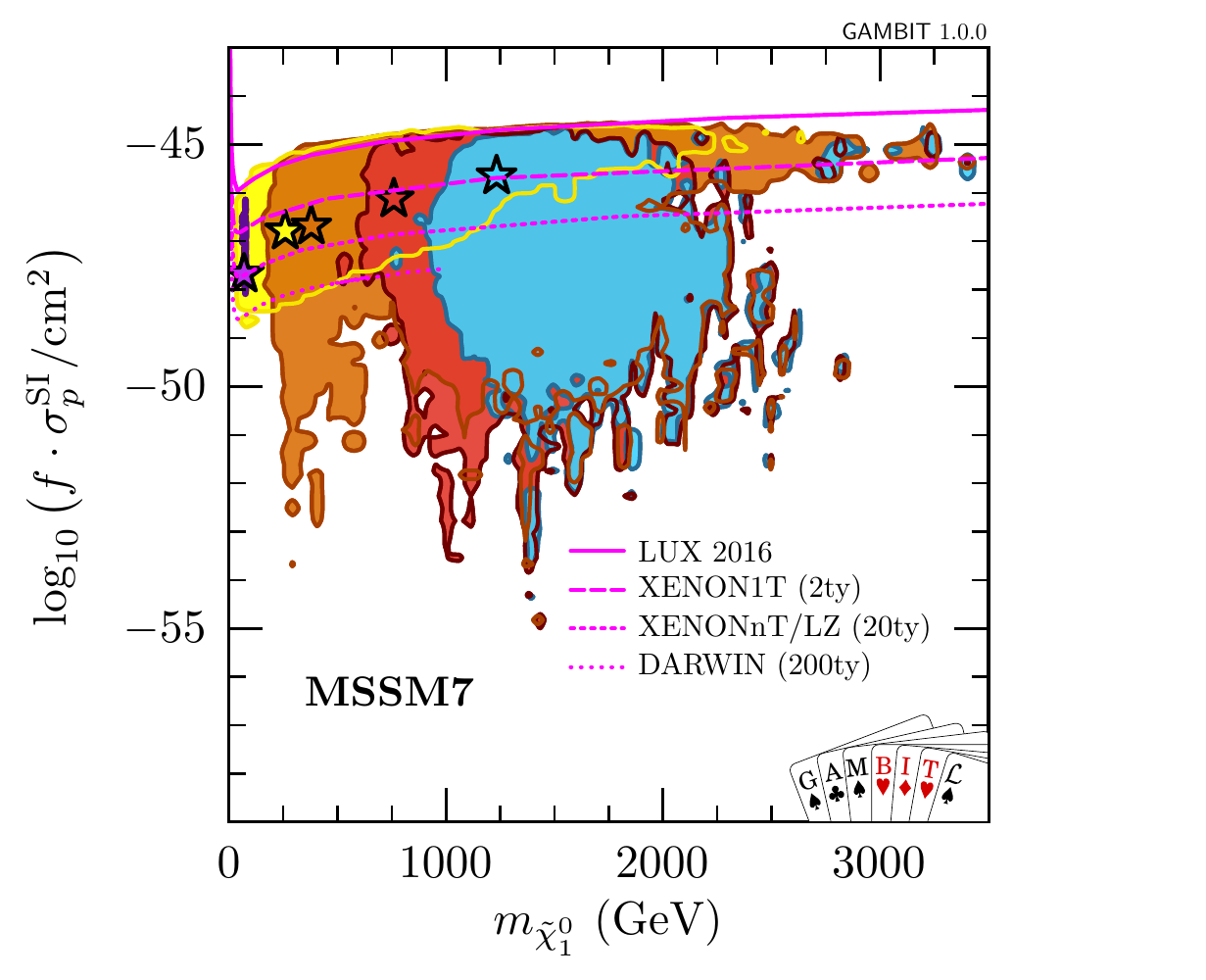}\\ 
\includegraphics[height=2.5in]{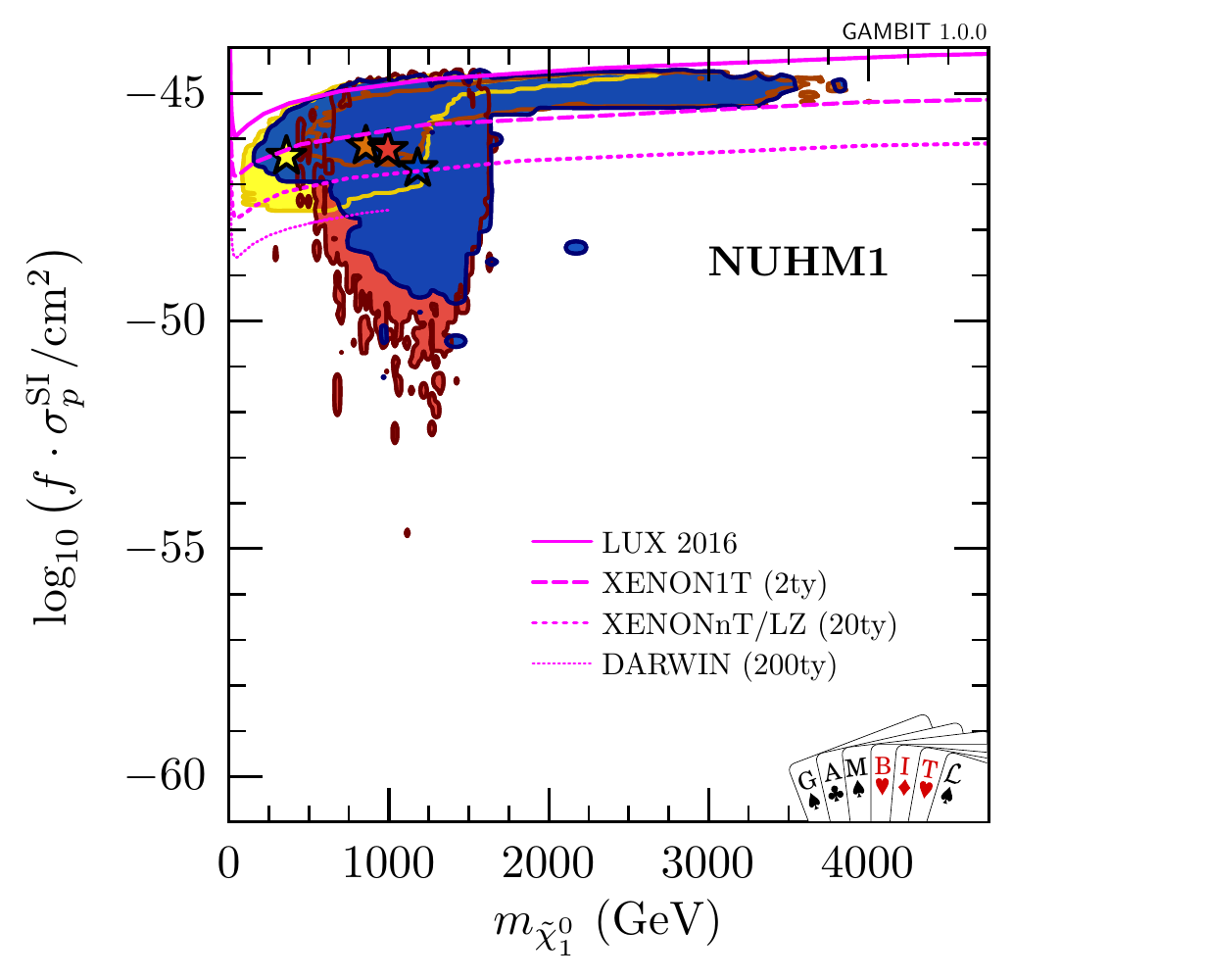}
\includegraphics[height=2.5in]{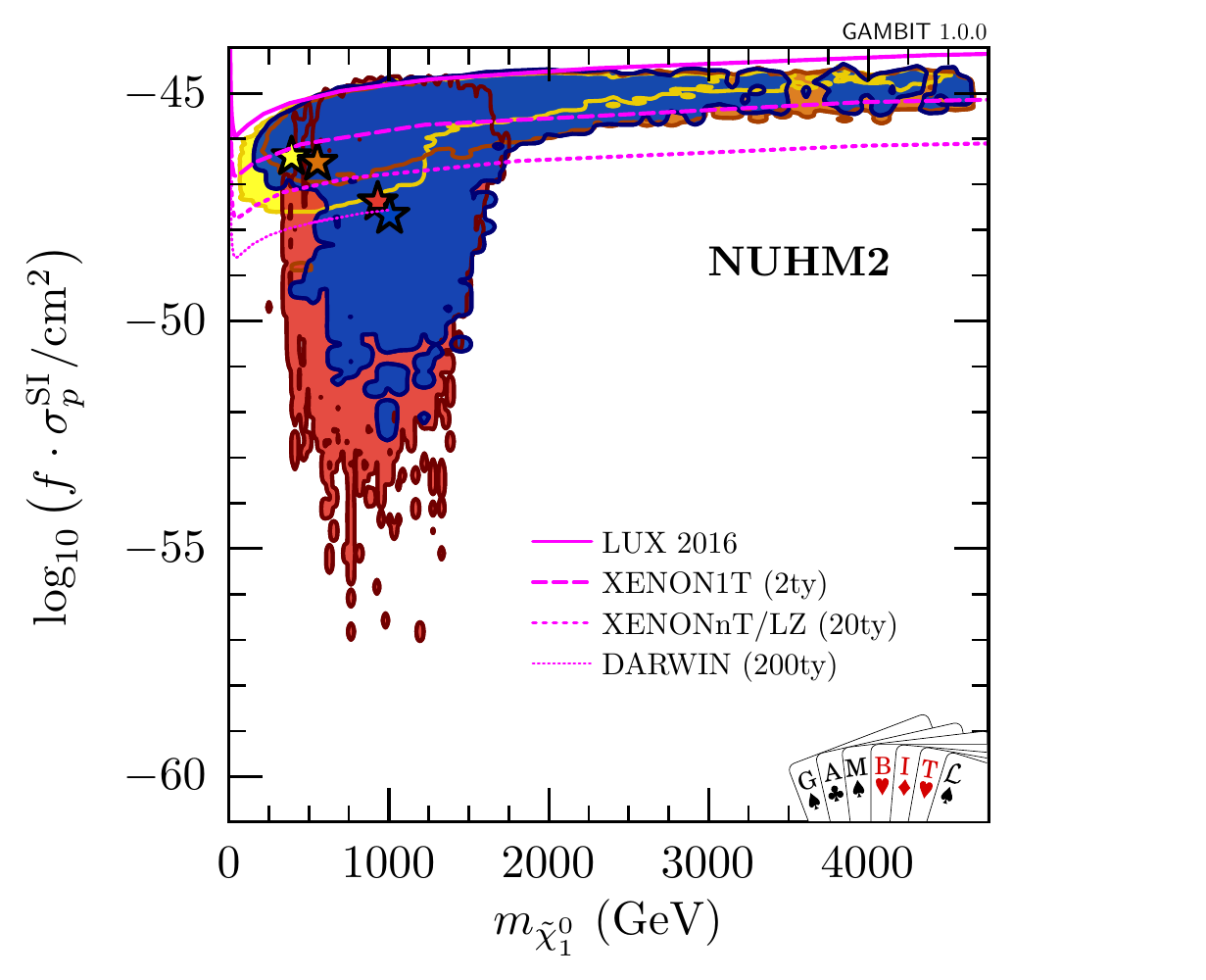}
\caption{ Spin independent neutralino-proton cross-sections for the
  CMSSM (top left), MSSM7 (top right), NUHM1 (bottom left) and NUHM2
  (bottom right) within our $2 \sigma$ confidence regions. The
  colours indicate active mechanism for depleting the relic density,
  with colours as in Fig.\ \ref{fig:NUHM} with sbottom co-annihilation
  (light blue) and $h/Z$ funnel (purple) now included as well for the
  MSSM7. The LUX 2016 limit (solid line) \cite{Akerib:2016vxi} with
  projections for limits from XENON1T (long dashed line), XENONnT and
  LZ (small dashed line) \cite{Schumann:2015cpa} and DARWIN (dotted
  line) \cite{Aalbers:2016jon} are all shown in pink.}
\label{fig:DDSI}
\end{figure}

The prediction for the spin independent neutralino-proton
cross-section for all three models discussed so far, and also a weak
scale MSSM7, are shown in Fig.\ \ref{fig:DDSI}.  We see that the
XENON1T, XENONnT, LZ and DARWIN experiments can test the entire CMSSM
chargino co-annihilation region.  This is the region with very heavy
sparticles that is challenging to detect at the LHC, so it means there
is a strong complementarity between different experiments.  On the
other hand the stop co-annihilation region is very challenging to
detect.  The situation is quite similar in the NUHM1, NUHM2 and MSSM7
models, though with the enlarged parameter space even the highly
efficient \GB scanning did not probe down to the extremely low
cross-sections that are seen in the CMSSM. In addition new regions such as the
stau co-annihilation (NUHM1 and NUHM2) and sbottom co-annihilation
(MSSM7) can also be seen extending to very low cross-sections in these models.





\section{Conclusions}
We have presented recent \GB results for MSSM models (CMSSM, NUHM1,
NUHM2 and MSSM7).  These reveal several different regions within the
$2 \sigma$ confidence interval.  In all models discussed there is a
chargino co-annihilation and heavy Higgs funnel region that is well
out if reach of the LHC, but can be probed by the forthcoming dark
matter direct detection experiments.  At the same time we see a
lighter stop co-annihilation region with better prospects for
discovery at the LHC and future colliders, but this region has spin
independent neutralino-proton cross-sections that are well below the
sensitivity of future dark matter detection experiments.  The stau
co-annihilation region is excluded from the $2 \sigma$ confidence
regions in the CMSSM and not present in the MSSM7, but is still
within $2 \sigma$ confidence limits in the NUHM1 and NUHM2.  In the
NUHM1 and NUHM2 the extra freedom also allows the chargino
co-annihilation and $A$ funnel regions to extend to lighter $m_0$ and
$m_{1/2}$ values, while in the MSSM7 we see new regions with sbottom
co-annihilation and $h/Z$ funnels open up.

\Acknowledgements
The work of PA was supported by Australian Research Council grants, FT160100274 and CE110001004.  PA gratefully acknowledges all members of the \GB collaboration, with whom this work was carried out.  

\bibliography{bibliography}

\begin{thebibliography}{99}



\bibitem{Athron:2017ard} 
  P.~Athron {\it et al.} [GAMBIT Collaboration],
  arXiv:1705.07908 [hep-ph].

\bibitem{Workgroup:2017htr} 
  G.~D.~Martinez {\it et al.} [GAMBIT Collaboration],
  arXiv:1705.07959 [hep-ph].


\bibitem{Workgroup:2017bkh} 
  P.~Athron {\it et al.} [GAMBIT Collaboration],
  arXiv:1705.07936 [hep-ph].

\bibitem{Workgroup:2017lvb} 
  T.~Bringmann {\it et al.} [GAMBIT Dark Matter Workgroup],
  arXiv:1705.07920 [hep-ph].

  
\bibitem{Balazs:2017moi} 
  C.~Bal\'azs {\it et al.} [GAMBIT Collaboration],
  arXiv:1705.07919 [hep-ph].

\bibitem{Workgroup:2017myk} 
  F.~U.~Bernlochner {\it et al.} [GAMBIT Collaboration],
  arXiv:1705.07933 [hep-ph].



\bibitem{Athron:2017kgt} 
  P.~Athron {\it et al.} [GAMBIT Collaboration],
  arXiv:1705.07931 [hep-ph].

  

\bibitem{Athron:2017qdc} 
  P.~Athron {\it et al.} [GAMBIT Collaboration],
  arXiv:1705.07935 [hep-ph].

\bibitem{Athron:2017yua} 
  P.~Athron {\it et al.} [GAMBIT Collaboration],
  arXiv:1705.07917 [hep-ph].


\bibitem{Feroz:2008xx} 
  F.~Feroz, M.~P.~Hobson and M.~Bridges,
  Mon.\ Not.\ Roy.\ Astron.\ Soc.\  {\bf 398}, 1601 (2009)

\bibitem{Athron:2014yba} 
  P.~Athron, J.~h.~Park, D.~St\"ockinger and A.~Voigt,
  Comput.\ Phys.\ Commun.\  {\bf 190}, 139 (2015)

\bibitem{Staub:2009bi} 
  F.~Staub,
  Comput.\ Phys.\ Commun.\  {\bf 181}, 1077 (2010)
  doi:10.1016/j.cpc.2010.01.011
  [arXiv:0909.2863 [hep-ph]].

\bibitem{Staub:2010jh} 
  F.~Staub,
  Comput.\ Phys.\ Commun.\  {\bf 182}, 808 (2011)
  doi:10.1016/j.cpc.2010.11.030
  [arXiv:1002.0840 [hep-ph]].


\bibitem{Staub:2012pb} 
  F.~Staub,
  Comput.\ Phys.\ Commun.\  {\bf 184}, 1792 (2013)
  doi:10.1016/j.cpc.2013.02.019
  [arXiv:1207.0906 [hep-ph]].


\bibitem{Staub:2013tta} 
  F.~Staub,
  Comput.\ Phys.\ Commun.\  {\bf 185}, 1773 (2014)
  doi:10.1016/j.cpc.2014.02.018
  [arXiv:1309.7223 [hep-ph]].

  
\bibitem{Allanach:2001kg} 
  B.~C.~Allanach,
  Comput.\ Phys.\ Commun.\  {\bf 143}, 305 (2002)
  doi:10.1016/S0010-4655(01)00460-X
  [hep-ph/0104145].


\bibitem{Allanach:2013kza} 
  B.~C.~Allanach, P.~Athron, L.~C.~Tunstall, A.~Voigt and A.~G.~Williams,
  Comput.\ Phys.\ Commun.\  {\bf 185}, 2322 (2014)
  doi:10.1016/j.cpc.2014.04.015
  [arXiv:1311.7659 [hep-ph]].

  

\bibitem{Djouadi:1997yw} 
  A.~Djouadi, J.~Kalinowski and M.~Spira,
  Comput.\ Phys.\ Commun.\  {\bf 108}, 56 (1998)

\bibitem{Muhlleitner:2003vg} 
  M.~Muhlleitner, A.~Djouadi and Y.~Mambrini,
  Comput.\ Phys.\ Commun.\  {\bf 168}, 46 (2005)

  
\bibitem{Djouadi:2006bz} 
  A.~Djouadi, M.~M.~Muhlleitner and M.~Spira,
  Acta Phys.\ Pol-on.\ B {\bf 38}, 635 (2007)

\bibitem{Bechtle:2008jh} 
  P.~Bechtle, O.~Brein, S.~Heinemeyer, G.~Weiglein and K.~E.~Williams,
  Comput.\ Phys.\ Commun.\  {\bf 181}, 138 (2010)

\bibitem{Bechtle:2013xfa} 
  P.~Bechtle, S.~Heinemeyer, O.~St{\aa}l, T.~Stefaniak and G.~Weiglein,
  Eur.\ Phys.\ J.\ C {\bf 74}, no. 2, 2711 (2014)

\bibitem{Sjostrand:2014zea} 
  T.~Sj\"ostrand {\it et al.},
  Comput.\ Phys.\ Commun.\  {\bf 191}, 159 (2015)

  
\bibitem{Athron:2015rva} 
  P.~Athron {\it et al.},
  Eur.\ Phys.\ J.\ C {\bf 76}, no. 2, 62 (2016)

  
\bibitem{Mahmoudi:2007vz} 
  F.~Mahmoudi,
  Comput.\ Phys.\ Commun.\  {\bf 178}, 745 (2008)

  
\bibitem{Belanger:2001fz} 
  G.~Belanger, F.~Boudjema, A.~Pukhov and A.~Semenov,
  Comput.\ Phys.\ Commun.\  {\bf 149}, 103 (2002)

\bibitem{Gondolo:2004sc} 
  P.~Gondolo, J.~Edsjo, P.~Ullio, L.~Bergstrom, M.~Schelke and E.~A.~Baltz,
  JCAP {\bf 0407}, 008 (2004)

\bibitem{Scott:2012mq} 
  P.~Scott {\it et al.} [IceCube Collaboration],
  JCAP {\bf 1211}, 057 (2012)

  
\bibitem{Han:2016gvr} 
  C.~Han, K.~i.~Hikasa, L.~Wu, J.~M.~Yang and Y.~Zhang,
  Phys.\ Lett.\ B {\bf 769}, 470 (2017)

\bibitem{Akerib:2016vxi} 
  D.~S.~Akerib {\it et al.} [LUX Collaboration],
  Phys.\ Rev.\ Lett.\  {\bf 118}, no. 2, 021303 (2017)

\bibitem{Schumann:2015cpa} 
  M.~Schumann, L.~Baudis, L.~B\"utikofer, A.~Kish and M.~Selvi,
  JCAP {\bf 1510}, no. 10, 016 (2015)

\bibitem{Aalbers:2016jon} 
  J.~Aalbers {\it et al.} [DARWIN Collaboration],
  JCAP {\bf 1611}, 017 (2016)
  
\end{thebibliography}

\end{document}